\documentclass[a4paper,10pt,twocolumn]{article}
\usepackage{amsmath}
\usepackage{amssymb}
\usepackage{graphicx}
\usepackage{float}
\usepackage[small]{caption}
\usepackage[a4paper, left=1.45cm, right=1.45cm, top=3.5cm, bottom=3.5cm, headsep=1.2cm]{geometry}
\usepackage{abstract}
\usepackage{hyperref}

\usepackage{authblk}

\title{The Sommerfeld enhancement for scalar particles and application to sfermion co-annihilation regions}
\author{Andrzej Hryczuk} 
\affil{\normalsize{SISSA and INFN - Sezione di Trieste} \\
{\small via Bonomea 265, 34136 Trieste, Italy}\\
{\small e-mail:hryczuk@sissa.it}}

\date{}

\begin{document}

\twocolumn[
\maketitle
\begin{onecolabstract}
{\normalsize
We study the impact of the Sommerfeld enhancement on the thermal relic density of the lightest neutralino in the case of large co-annihilation effects with a scalar particle. The proper way of including the Sommerfeld effect in this case is discussed, and the appropriate formulas for a general scenario with a set of particles with arbitrary masses and (off-)diagonal interactions are provided. We implement these results to compute the relic density in the neutralino sfermion co-annihilation regions in the mSUGRA framework. We find non-negligible effects in whole sfermion co-annihilation regimes. For stau co-annihilations the correction to the relic density is of the order of several per cent, while for stop co-annihilations is much larger, reaching a factor of 5 in some regions of the parameter space. A numerical package for computing the neutralino relic density including the Sommerfeld effect in a general MSSM setup is made public available.
}
\end{onecolabstract}
\vspace{1cm}
]

\section{Introduction}
\label{sec:intro}

In recent years, the existence of dark matter (DM) has been well established and its density measured with a few per cent level of precision. A recent analysis, within the 6-parameter $\Lambda$CDM model, of the 7-year 
WMAP data~\cite{Jarosik:2010iu}, combined with the Baryon Acoustic Oscillations ~\cite{Percival:2009xn} and the recent redetermination of the Hubble constant ~\cite{Riess:2009pu}, 
gives \cite{Komatsu:2010fb}: $\Omega_{\rm{CDM}} h^2 = 0.1123\pm0.0035$, where $\Omega$ is the ratio between mean density and critical density, and $h$ is the Hubble constant in units of $100\,{\rm km}\,{\rm s}^{-1}\,{\rm Mpc}^{-1}$.

One of the most attractive scenarios explaining the nature of dark matter is that it is composed of stable weakly interacting massive particles (WIMPs), since their thermal relic abundance is naturally of the order of the measured one. However, in order to predict its precise value for a given model a careful calculation has to be done. In particular, there are several effects that can alter the relic density of a thermal relic and which need to be taken into account.

In this Letter we contribute to the study of one of these effects, i.e. the Sommerfeld enhancement \cite{Sommerfeld}. It is a non-relativistic effect changing the annihilation cross section due to a long range force acting between slowly moving initial particles. This effect has been studied widely recently (see e.g. \cite{Hryczuk:2010zi} and references therein). Although its applications were discussed for both scalar and fermion initial states, the derivation of Sommerfeld corrections for a general multi-state case was given explicitly only for fermions \cite{Iengo:2009ni} (see also \cite{Cassel:2009wt} for a different approaches). Here we extend it to the case of scalar-scalar and fermion-scalar pairs and discuss possible applications. Finally, we present some numerical results for the impact of the Sommerfeld effect on the relic density of the neutralino in stau and stop co-annihilation regions in minimal supergravity (mSUGRA) scenario.

\section{Sommerfeld enhancement for scalars}
\label{sec:analitic}

We consider an annihilation process of two particles, $\varphi_{i}$~and $\varphi_{j}$, which are coupled to some light interaction boson $\phi$, leading to a long range interaction between them. In the case when this interaction is diagonal (i.e. the exchange of $\phi$ does not change the particles), in the non-relativistic limit the spin of initial particles does not matter - the static force is the same for both scalars and fermions. This is however not true if interactions can be off-diagonal and intermediate particles can have different masses. Then due to the differences in the couplings and propagators between scalars and fermions, the computations of the Sommerfeld effect slightly differ.
\par 
In deriving the effect for this case we follow the approach of Ref. \cite{Iengo:2009ni}, where a general method of computing the Sommerfeld enhancement from the field theory diagrams was presented: to obtain the Sommerfeld enhancement factors $S_{ij}$ one has to solve the set of Schr\"{o}dinger equations for the two-body wave-functions $\psi_{ij}$:
\begin{eqnarray}
\label{schrr}
 -\frac{\partial^2}{2m_r^{ij}}\psi_{ij} (\vec{r})= &&U_{ij}^0(\vec{r})+\left(\mathcal{E}-2\delta m_{ij} \right)\psi_{ij} (\vec{r})\nonumber\\
&+&\sum_{i'j'\phi}V^\phi_{ij,i'j'}\psi_{i'j'}(\vec{r}),
\end{eqnarray}
and then compute
\begin{equation}
 S_{ij}=\left|\frac{\psi_{ij}(\infty)}{\psi_{ij}(0)}\right|^2,
\end{equation} 
where $m_r^{ij}$ is the reduced mass of annihilating $\varphi_{i}\varphi_{j}$ pair, $U_{ij}^0$ contains the tree-level amplitude and the sum is over (possibly different) $\varphi_{i'}\varphi_{j'}$ intermediate states and different interactions. Here $\mathcal{E}~=~{\vec{p}}\,^2/2m_r^{ab}$ is the kinetic energy of the incoming pair (at infinity), with $m_r^{ab}$ its reduced mass and $\vec{p}$ the CM three-momentum; $2\delta m_{ij}=m_{i}+m_{j}-(m_a+m_b)$ is the mass splitting (for more details see Ref. \cite{Hryczuk:2010zi}). The potential has the form:
\begin{equation}
\label{potentialr}
 V^\phi_{ij,i'j'}(r)= \frac{c_{ij,i'j'}(\phi)}{4\pi}\frac{e^{-m_\phi r}}{r}\, ,
\end{equation}
where $c_{ij,i'j'}(\phi)$ are coefficients depending on the couplings and states involved. For incoming fermions the coefficients were presented in Tab. 1 of Ref. \cite{Hryczuk:2010zi}. Below we give results for scalar-scalar and fermion-scalar pairs.
\begin{figure}
 \centering
 \includegraphics[scale=0.32]{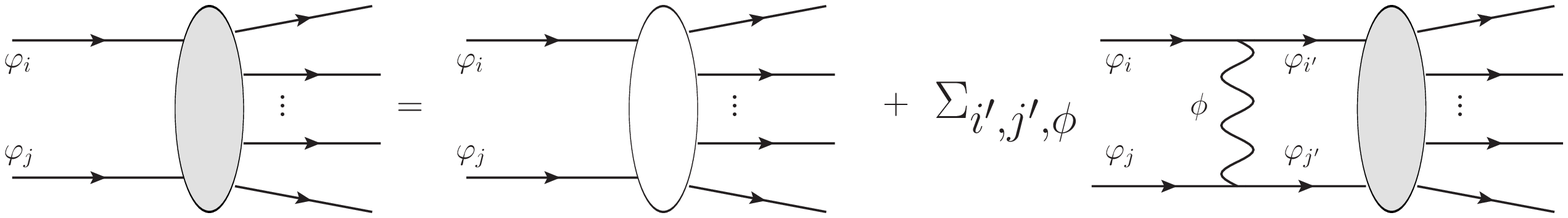}
 \caption{Recursive relation for the full annihilation amplitude including the Sommerfeld enhancement. Blob represents any possible annihilation process.}
 \label{fig:diagram}
\end{figure}

To obtain those coefficients we use a method which is a straightforward generalization of the one developed in Ref.~\cite{Iengo:2009ni}. We write a recurrence relation for the annihilation amplitudes as visualized in Fig.~\ref{fig:diagram}. Assuming that $\delta m_{ij}/m_i\ll 1$, in the non-relativistic limit we can transform this integral equation to the Schr\"{o}dinger one, obtaining automatically the coefficients $c_{ij,i'j'}(\phi)$ in the potential.
\par
Let's denote by a superscript $S$ the case with two scalars and by $F$ with one scalar and one fermion. In the second case let $i$ and $i'$ be fermions and $j$ and $j'$ scalars. Then we find:
\begin{equation}
 c_{ij,i'j'}=g^\phi_{ii'}g^\phi_{jj'} N^{S,F}_{ij,i'j'} A^{S,F}_\phi(m_i,m_j,m_{i'},m_{j'}) \, ,
\end{equation} 
where $g^\phi_{ii'}$ is a coupling present in the $ii'\phi$ vertex; the normalization and combinatorics gives
\begin{eqnarray*}
 &&N^S_{ij,i'j'}=\Biggl\lbrace 
	\begin{array}{ll}
	1 & \quad i=j, i'=j'\ \rm{or}\ i\neq j, i'\neq j' , \cr
	\sqrt{2} & \quad i\neq j, i'=j' \ \rm{or}\ i= j, i'\neq j',
	\end{array} \\
 &&N^F_{ij,i'j'}=1 \, ,
\end{eqnarray*} 

and factors $A^{S,F}_\phi$ are, with $\phi=V,A,S$ indicating respectively a vector, an axial vector and a scalar:
\begin{eqnarray}
\label{ASV}
 A^S_V&=&A^S_A= \frac{1}{2}\left(1+\frac{m_i}{2m_{i'}}+\frac{m_j}{2m_{j'}}\right), \\
\label{ASS}
 A^S_S&=&\frac{1}{4m_{i'} m_{j'}}, \\
\label{AFV}
 A^F_V&=&\frac{m_{j'}+m_j}{2m_{j'}}\, , \\
\label{AFA}
 A^F_A&=&0 \, ,\\
\label{AFS}
 A^F_S&=&\frac{1}{2m_{j'}}.
\end{eqnarray}
In the limit when all the masses are equal coefficients $A^S_V,\ A^S_A$ reduce to the ones which were used in Refs. \cite{Freitas:2007sa,Cirelli:2007xd,Berger:2008ti}. However, in general case when the masses of intermediate scalars differ, Eqs. (\ref{ASV}) and (\ref{ASS}) have to be used. Neglecting this fact can give rise to several percent difference in the $c_{ij,i'j'}(\phi)$ coefficients.\footnote{Some of these coefficients are divergent in the limit when one of the masses vanish. However, in this case the non-relativistic approximation does not hold and hence these results are not valid.}  We stress once again that those results are valid in the non-relativistic limit and when mass splitting is much smaller than all of the masses involved.
\par
All the considerations above implicitly assumed that the interaction strength is sufficiently weak that the higher loop corrections do not alter the potential significantly. This is true for the weak and electromagnetic interactions, as well as for the Higgs exchange. However, in the case of strong interactions, corrections to the gluon exchange coming from gluon self-interactions and fermion loops may become important. To take this into account, following \cite{Freitas:2007sa}, instead of the potential (\ref{potentialr}) we will use one computed in \cite{Schroder:1998vy}, which in the configuration space is:\footnote{Note that here the interaction is diagonal and one does not need to use coefficients (\ref{ASV}-\ref{AFS}).}
\begin{eqnarray}
\label{QCDpot}
 V(\vec{r})=&-&C_F\frac{\alpha_s}{r}-C_F\frac{\alpha_s^2}{4\pi}\frac{1}{r}\Biggl[\frac{31}{9}C_A-\frac{20}{9}T_F n_f\nonumber\\
&+&\beta_0\bigl(2\gamma_E+\log (\mu^2 r^2)\bigr)\Biggr]+\mathcal{O}(\alpha_s^3),
\end{eqnarray}
where $\beta_0=\frac{11}{3} C_A-\frac{4}{3}T_F n_f$, $n_f$ is the number of massless quarks (we choose it to be 5, since the stop co-annihilation is most important in the $\mathcal{O}$(100~GeV) region, where the top mass $m_t$ is non negligible) and Euler gamma is $\gamma_E\approx0.5772$. For the case of SU(3) we have $C_F=4/3$, $T_F=1/2$, $C_A=3$. For the QCD scale we take $\mu^2= 2m_t^2$.
\par
Another effect one has to take into account is the presence of thermal corrections, as first discussed in a similar context in \cite{Cirelli:2007xd}. They change the exchanged boson masses and in particular photon and gluon become massive, which introduces a Yukawa cut-off to the potential in Eq. (\ref{QCDpot}). These corrections may be also important for the mass splittings, if there are nearly degenerate states present in the spectrum.

There are two types of thermal effects which we need to include as discussed in \cite{Cirelli:2007xd,Hryczuk:2010zi}: the scaling of the Higgs VEV with the temperature  \cite{Dine:1992wr}, $v(T)=v {\rm Re} \sqrt{1-T^2/T^2_c}$ where we took $T_c=200$ GeV, and the contribution to gauge boson masses due to the screening by the thermal plasma; the so-called Debye mass~\cite{Gross:1980br}. For the gluon the screening of the plasma introduces at a leading order a contribution \cite{Kajantie:1997pd}:
\begin{equation}
 m^2_g=(N_c/3+N_f/6)g_s^2T^2=\frac{3}{2}g_s^2T^2.
\end{equation}

\section{Applications}
\label{sec:appl}

To have an idea when the Sommerfeld effect can have a non-negligible impact on the relic density it is useful to give some approximate general conditions, which have to be satisfied by the dark matter particle or the co-annihilating one: i) coupling to the boson with much lower mass (``long range force''), ii) the coupling strength at least of the order of the weak coupling and iii) if the effect comes from the co-annihilating particle, small mass splitting between it and the DM.
\par
Those conditions are not easy to satisfy without invoking some new interactions. The reason is that since the DM has to be electromagnetically neutral, it can couple only to $Z$, $W^\pm$ and Higgs bosons, all of which are heavy. This pushes up the region of possible Sommerfeld effect influence to very large masses. Hence, without additional interactions, the only possibility is the impact of co-annihilating particles, which can have very different quantum numbers, i.e. they can even have both electromagnetic and color charge. In this case their annihilation cross sections can be altered significantly by the Sommerfeld effect coming form the exchange of photons and/or gluons. If such co-annihilating particle is degenerate with the DM, the total effective annihilation cross section can get large corrections.
\par
The co-annihilation regime is typically the only place where Sommerfeld effect can be significant in the minimal supersymmetric standard model (MSSM), or in any other theory beyond the Standard Model without new interactions. We will concentrate on applications to mSUGRA, in which the dark matter candidate is the lightest neutralino $\tilde\chi^0_1$ with mass $m_\chi$. This is because in this Letter we are interested in the effect for scalar particles and in this framework there are two extensively studied parameter regions involving co-annihilation with a scalar particle.\footnote{In the MSSM there are other well motivated cases possible: the chargino co-annihilation region discussed in \cite{Hryczuk:2010zi,Hisano:2004ds}, sneutrino DM scenarios \cite{Falk:1994es}  (however in some tension with experiment) and possibly some small effect could be found in the ``well tempered'' neutralino \cite{ArkaniHamed:2006mb} in the scenarios in which Higgs is light. Note that in less motivated cases, eg. when SUSY breaking scale is very high or when several states are degenerate with $\tilde\chi^0_1$, the Sommerfeld effect gives very large corrections and is essential for reliable relic density calculation.} Those are the stau $\tilde\tau$ \cite{Ellis:1998kh} and stop $\tilde t$ co-annihilation regions \cite{Boehm:1999bj}.
\par
The existence of the Sommerfeld effect in the stau case was first suggested by authors of Ref. \cite{Hisano:2006nn} (a one-loop manifestation of this effect was also discussed in \cite{Baro:2007em}), where they pointed out that although $\tilde \tau^+ \tilde \tau^-$ annihilation exhibits a strong enhancement, the $\tilde \tau^\pm \tilde \tau^\pm$ are strongly suppressed, and that the net result should not be very large. Here we will explicitly show with full numerical calculations that this is indeed the case, and discuss in mSUGRA its strong dependence on the value of $\tan\beta$. In the stop co-annihilation region this effect was first discussed by Freitas \cite{Freitas:2007sa}, where the QCD corrections to the bino-stop co-annihilations were considered, among which the Sommerfeld one was dominant. Below we will show our results also for this case.\footnote{Our setup is slightly more general, because we do not make any assumption on the neutralino composition. This however do not introduce any significant difference in the result, since in the stop co-annihilation regime in mSUGRA the Wino and Higgsino component in the lightest neutralino is very small.}
\par
In order to compute the relic density, we have written a numerical code and implemented it into DarkSUSY \cite{Gondolo:2004sc}, which allows to compute the relic density with 1\% accuracy.

\paragraph{Stau co-annihilation} 

\begin{figure}[t!]
 \centering
 \includegraphics[scale=0.7]{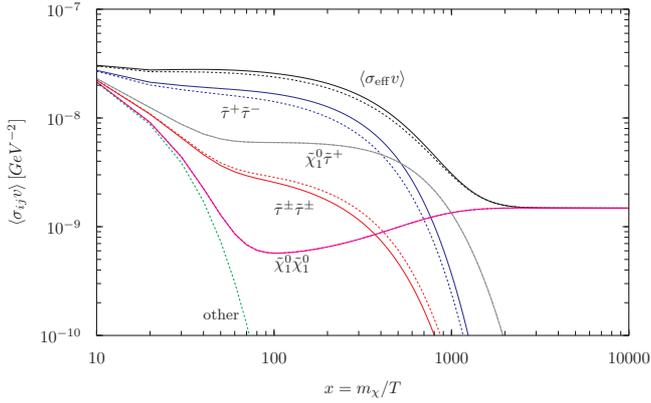}
 \caption{An example of the thermally averaged effective annihilation cross section, for the stau co-annihilation region. Contributions coming from different annihilation processes are indicated. The solid lines correspond to the case with Sommerfeld correction included, while dashed ones without it. In the $\tilde\chi^0_1\tilde\chi^0_1$ case the effect is too small to be visible. Parameters are: $m_\chi=182.7$ GeV, $m_{\tilde\tau}=183.2$ GeV, $\tan\beta=50$, $A_0=0$ and $\mu>0$.}
 \label{fig:sigmaeff_chn}
\end{figure}
In this case the impact of the Sommerfeld effect is relatively mild. The reason is that, although at the freeze-out temperature the thermally averaged annihilation cross section $\langle \sigma_{\rm eff}v\rangle$ is dominated by the contribution coming from $\tilde\tau\tilde\tau$ annihilation, the net enhancement is rather small, since as can be seen in Fig.~\ref{fig:sigmaeff_chn} there are both attractive and repulsive modes present. 

\begin{figure}
 \centering
 \includegraphics[scale=0.81]{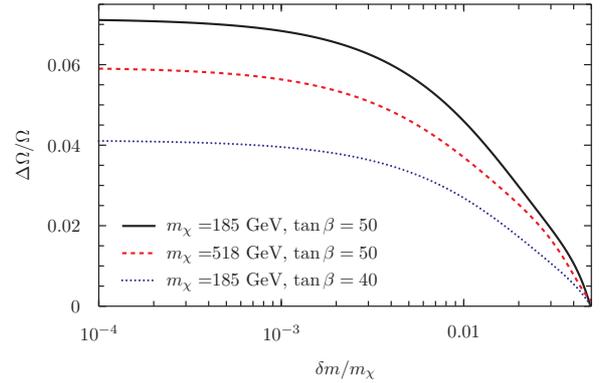}
 \caption{The correction to the relic density $\Delta\Omega=\Omega_0-\Omega_{SE}$ coming from the Sommerfeld effect. Results for three different parameter points are given to show that the result is much more sensitive to $\tan\beta$ than $m_\chi$. Note that the accuracy of the relic density computation in DarkSUSY is about 1\%.}
 \label{fig:enh}
\end{figure}

\begin{figure}
 \centering
 \includegraphics[scale=0.825]{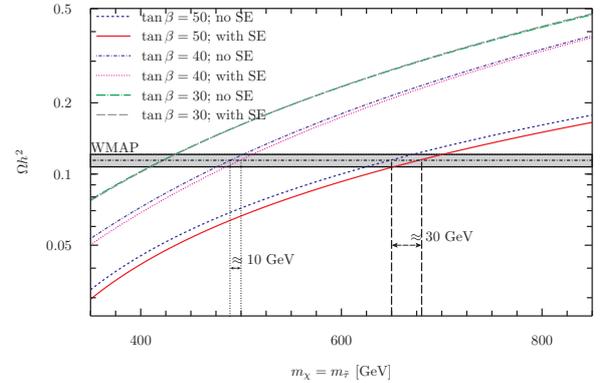}
 \caption{Effect of the Sommerfeld enhancement on the relic density for $m_\chi=m_{\tilde\tau}$. Maximal neutralino mass giving relic density compatible with data, and its shift due to the Sommerfeld effect are highlighted.}
 \label{fig:oh2diag}
\end{figure}

Furthermore, if one enlarges the mass splitting, the co-annihilations become less effective and the total effect on the relic density gets even smaller. This has been shown in Fig.~\ref{fig:enh}, from which one can see that the Sommerfeld effect introduces a several per cent correction to the relic density, but only in the region of a very degenerate $\tilde\tau$. This region in most of the parameter space does not give the relic density compatible with the WMAP data. However, the impact of the Sommerfeld enhancement become more important when $m_\chi$ grows, because then the WMAP contour approaches the region where $m_\chi\simeq m_{\tilde \tau}$: the $\tilde\chi^0_1\tilde\chi^0_1$ annihilation cross section scales as $\sim m_\chi^{-2}$ and to get the same $\langle \sigma_{\rm eff}v\rangle$ (and the same relic density) one has to compensate with larger co-annihilation effects. This means, in particular, that the maximal $\tilde\chi^0_1$ mass that can give correct relic density gets shifted by a Sommerfeld effect by a sizable amount. In mSUGRA the value of this shift depends strongly on $\tan\beta$. In Fig.~\ref{fig:oh2diag} we show the dependence of $\Omega h^2$ vs. the neutralino mass in the case where it is equal to the stau mass for its three different values. The maximal effect is seen for large $\tan\beta$ and drops down quite considerably when it is decreased. The reason for this is that for higher $\tan\beta$ the $\tilde\tau^+\tilde\tau^-\rightarrow h^0 h^0$ annihilation becomes very efficient. On the other hand the annihilation cross sections of $\tilde\tau^\pm\tilde\tau^\pm$ to two leptons or quarks also grow with $\tan\beta$ but much slower, hence enlarging its value makes attractive channels dominant over repulsive ones (as in the case in Fig.~\ref{fig:sigmaeff_chn}).

\paragraph{Stop co-annihilation} 

In mSUGRA, in the region of parameter space where the large negative value of $A_0$ drives the lighter stop to be degenerate with the neutralino, the results of the relic density computations are significantly affected by the Sommerfeld effect. 

\begin{figure}
 \centering
 \includegraphics[scale=0.64]{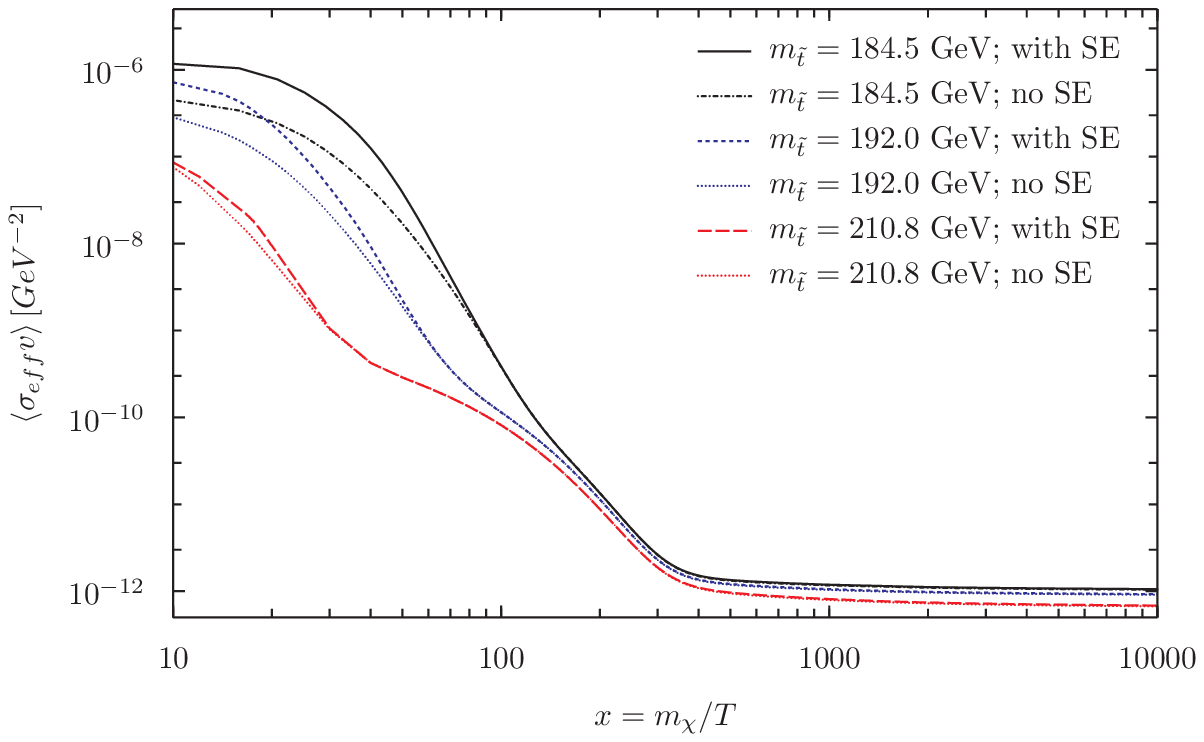}
 \caption{An example of the thermally averaged effective annihilation cross section for the stop co-annihilation region. Parameters are $m_\chi=140.7$ GeV, $\tan\beta=10$, $A_0=-2750$ GeV and $\mu>0$. Two thresholds are visible, smeared by the thermal average, $\tilde\chi^0_1 \tilde t$ and $\tilde t \tilde t$, but only the second one gets significant Sommerfeld correction.}
 \label{fig:sigmaeff_stop}
\end{figure}

Fig. \ref{fig:sigmaeff_stop} shows the change of the thermal averaged effective annihilation cross section due to the Sommerfeld effect for different mass splittings between $\tilde\chi^0_1$ and $\tilde t$. The correction is significantly larger than in the $\tilde\tau$ case, because of the strong force coming from the gluon exchange. When the mass splitting becomes larger two effects can be seen. Firstly, the $\tilde t \tilde t$ threshold occurs for higher temperature, which lowers the overall impact of the co-annihilating particle. Secondly, also the magnitude of the correction to $\langle \sigma_{\rm eff}v\rangle$ becomes smaller, since at higher $T$ the typical velocities are higher and moreover, thermal corrections to the gluon mass are larger.

\begin{figure}
 \centering
 \includegraphics[scale=0.825]{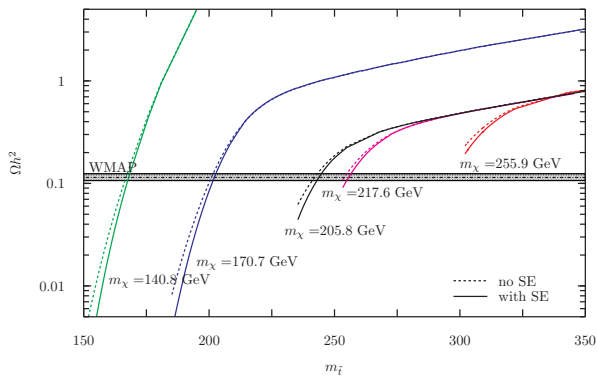}
 \caption{Effect of the Sommerfeld enhancement on the relic density for the stop co-annihilation region. For given $m_\chi$ there is a lower bound on $m_{\tilde t}$ due to the constraint on the lightest Higgs mass.}
 \label{fig:oh2stop}
\end{figure}

This change in $\langle \sigma_{\rm eff}v\rangle$ can affect considerably the relic density of the neutralino. The results for $\Omega h^2$ with and without the Sommerfeld effect included are presented in Fig.~\ref{fig:oh2stop} for five different $m_\chi$. One can see that the largest effect is obtained for parameters giving typically too small relic density. Nevertheless, in the region compatible with WMAP results, the correction can still be larger than the current observational uncertainty. 

\begin{figure}
 \centering
 \includegraphics[scale=0.8]{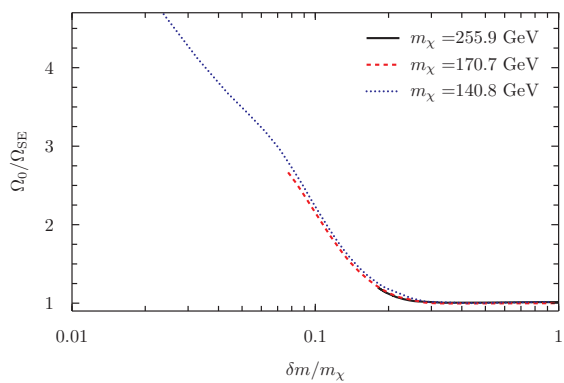}
 \caption{Sommerfeld effect on the relic density for $\tilde t$ very degenerate with $\tilde\chi^0_1$. As before, due to the bound on the lightest Higgs mass the degeneracy is limited; the more the larger $m_\chi$.}
 \label{fig:enh_stop}
\end{figure}

The importance of the Sommerfeld effect itself is more clearly seen in the Fig.~\ref{fig:enh_stop}, where the ratio of relic densities without and with SE is presented. For very degenerate stops $\Omega h^2$ can be suppressed by a factor of few.
\par
The results we presented in this subsection are in qualitative agreement with those in \cite{Freitas:2007sa}. However, there are slight quantitative differences, for several reasons. First of all, in this work we were interested in the Sommerfeld effect and we did not compute other QCD corrections. On the other hand, our treatment of the Sommerfeld enhancement is more accurate, since we include not only gluon exchange, but all possible interactions, and for all annihilation processes, not only for $\tilde t\tilde t$ one. We include also thermal corrections which modify the masses of exchanged bosons.
\par
Finally, we would like to point out that the results in the stop co-annihilation region are subject to sizable theoretical uncertainties. The reason is that since the coupling is relatively strong, Sommerfeld enhancement factors differ considerably from 1 even at high velocities. This cannot be however the true result, since the full quantum field theory initial state corrections in this case are not expected to be large. This discrepancy comes from the fact that the formalism used to compute the Sommerfeld enhancement is not valid in this regime. In our numerical calculations we used an approach to approximate the true corrections by the non-relativistic ones normalized in such a way that they vanish for $v\rightarrow 1$ (a better approximation would be to compute the NLO vertex correction, which is however beyond the scope of this work).\footnote{The normalization was done additively, i.e. the enhancement factors were shifted by a small constant value (always less than 1). Since for small velocities the enhancement is much larger than 1, in the non-relativistic regime this procedure does not introduce any significant change. This approach gives the relic density larger by at most 10\% with respect to the case with the Sommerfeld factors simply extrapolated to the high velocities regime.} To obtain more reliable predictions for the intermediate regime of velocities $\mathcal{O}(10^{-1})$, which are very important for precise relic density computation, one should refine the theoretical calculations beyond the non-relativistic techniques used to derive the Sommerfeld enhancement.

\section{Conclusions}
\label{sec:conlc}

In this Letter we studied the Sommerfeld effect for the scalar-scalar and fermion-scalar pairs. We gave general results for the coefficients in the interaction potential used to compute the enhancement, which to the extent of our knowledge were not discussed in the literature before. Those coefficients have to be used if more than one type of particle is present in the computation of the Sommerfeld effect for a setup with scalar particles. Although, most applications involve only one state, there are several interesting cases in which presented results may lead to modified phenomenological implications. Among these are: a general MSSM setup with heavy neutralino, $m_\chi\gtrsim 1$ TeV, degenerated with one (or more) sfermion, or sneutrino DM scenarios, again if it is degenerated with on or more sleptons. It is also worth to note that new possibilities open up if one goes beyond the MSSM, for instance in the next to minimal NMSSM (see eg. \cite{Ellwanger:2009dp} for a review). In particular, in the BMSSM framework \cite{Dine:2007xi} one can considerably lower the constraints on the lightest Higgs mass, which would give rise to large effects coming from the light Higgs exchange (e.g. to $\tilde\chi^0_1 \tilde t$ but even also to $\tilde\chi^0_1 \tilde\chi^0_1$ annihilation). 
\par
In the second part of the paper, we presented numerical results for the influence of this effect on the relic density of neutralino in the mSUGRA stau and stop co-annihilation regions. For the stop case we confirm and give some improvements over results of Ref. \cite{Freitas:2007sa}. In the stau case we get new results saying that the effect on the relic density ranges from about 1 to about 7\%. Typically this does not introduce sizable change of the $m_\chi$ which gives the correct relic density, however this change can be significant for the maximal mass, i.e. when $\tilde\chi^0_1$ is degenerated with $\tilde\tau$, giving rise to about 5\% shift. We observe that this effect on the maximal mass is largest for $\tan\beta\approx 50$, since then the relative contribution to the effective cross section of $\tilde\tau^+\tilde\tau^-$ vs. $\tilde\tau^\pm\tilde\tau^\pm$ annihilation is the largest, and becomes smaller when $\tan\beta$ is decreased.
\par
For the numerical computations we have developed a package for DarkSUSY, which is able to compute the Sommerfeld effect and its impact on the neutralino relic density for a general MSSM setup. We provide it as a public available tool\footnote{The package is available for download from the webpage: \url{http://people.sissa.it/~hryczuk}} to be used with DarkSUSY for obtaining reliable relic density predictions in all possible cases where this effect is relevant.

\paragraph{Acknowledgments}

The author would like to thank P.~Ullio and A.~Freitas for helpful discussions and to R.~Iengo and P.~Ullio for careful reading of the manuscript.

{\linespread{0.1}

\end{document}